\documentclass[english,notitlepage,reprint,unsortedaddress]{revtex4-2}
\usepackage[T1]{fontenc}
\usepackage[latin9]{inputenc}
\usepackage{textcomp}
\usepackage{mathtools}
\usepackage{amsmath}
\usepackage{amssymb}
\usepackage{mathdots}
\usepackage{graphicx}
\PassOptionsToPackage{version=3}{mhchem}
\usepackage{mhchem}

\makeatletter

\usepackage{babel}

\makeatother

\usepackage{babel}
\begin{document}
\title{Stochastic density functional theory combined with Langevin dynamics
for warm dense matter}
\author{Rebecca Efrat Hadad}
\affiliation{Fritz Haber Research Center for Molecular Dynamics, Institute of Chemistry,
The Hebrew University of Jerusalem, 91904, Israel}
\author{Argha Roy}
\address{Institute for Physics, University of Rostock, 18051 Rostock, Germany}
\author{Eran Rabani}
\affiliation{Department of Chemistry, University of California, Berkeley, California
94720, USA}
\affiliation{Materials Sciences Division, Lawrence Berkeley National Laboratory,
Berkeley, California 94720, USA }
\affiliation{The Raymond and Beverly Sackler Center of Computational Molecular
and Materials Science, Tel Aviv University, Tel Aviv 69978, Israel}
\author{Ronald Redmer}
\address{Institute for Physics, University of Rostock, 18051 Rostock, Germany}
\author{Roi Baer}
\affiliation{Fritz Haber Research Center for Molecular Dynamics, Institute of Chemistry,
The Hebrew University of Jerusalem, 91904, Israel}
\begin{abstract}
This study overviews and extends a recently developed stochastic finite-temperature
Kohn-Sham density functional theory to study warm dense matter using
Langevin dynamics, specifically under periodic boundary conditions.
The method's algorithmic complexity exhibits nearly linear scaling
with system size and is inversely proportional to the temperature.
Additionally, a novel linear-scaling stochastic approach is introduced
to assess the Kubo-Greenwood conductivity, demonstrating exceptional
stability for DC conductivity. Utilizing the developed tools, we investigate
the equation of state, radial distribution, and electronic conductivity
of Hydrogen at a temperature of 30,000K. As for the radial distribution
functions, we reveal a transition of Hydrogen from gas-like to liquid-like
behavior as its density exceeds 4 g/cm\textthreesuperior . As for
the electronic conductivity as a function of the density, we identified
a remarkable isosbestic point at frequencies around 7eV, which may
be an additional signature of a gas-liquid transition in Hydrogen
at 30,000K. 
\end{abstract}
\maketitle

\section{Introduction}

Warm dense matter (WDM) exists in the interior of planets \citep{guillot1999interiors,nguyen2004melting,redmer2011thephase,pozzo2012thermal,benuzzi-mounaix2014progress,french2019viscosity,ehrenreich2020nightside,fortney2010theinterior,bethkenhagen2017planetary,french2012textitab}
and in brown dwarfs \citep{saumon1992therole,kulkarni1997browndwarfs,becker2014abinitio,booth2015laboratory,becker2018material}
and white dwarf stars \citep{chabrier1993quantum,saumon2022current}.
In inertial and fusion systems \citep{hinkel2013scientific,lindl1995development},
WDM is generated by subjecting materials to high-energy lasers \citep{campbell2017laserdirectdrive}.
Despite its significance, exploring the diverse forms and compositions
of WDM poses a formidable challenge due to experimental complexities
associated with preparing and sustaining materials under extreme conditions
\citep{remington2005accessing,benuzzi-mounaix2014progress,falk2018experimental}.
Consequently, computational methods have become indispensable for
determining the equations of state as well as the chemical and physical
properties of different systems.

Among these methods are ab initio molecular dynamics (AIMD) calculations
\citep{militzer2001calculation,holst2008thermophysical,morales2010equation,brown2014quantum,bethkenhagen2020carbonionization},
which rely on an \textquotedbl adiabatic\textquotedbl{} approximation.
This approximation assumes that the quantum mechanical electrons reach
thermal equilibrium under the applied temperature, electronic chemical
potential, and Coulomb potentials corresponding to the instantaneous
positions of the atomic nuclei. The latter then undergo classical
motion under the conservative force derived from the electronic free
energy, effectively acting as a potential of mean force (see, e.g.,
ref. \citep{tuckerman2010statistical}). The observables can be determined
by averaging over a long adiabatic molecular dynamics (MD) trajectory.

The adiabatic and classical approximations behind AIMD have their
limitations. There is evidence suggesting that a classical approximation
may lack sufficient accuracy, particularly for temperatures below
1000K and under high pressures \citep{niu2023stablesolid,pickard2007structure}.
Moreover, nonadiabatic effects, though often disregarded, as done
here, have not been thoroughly explored in the context of AIMD in
WDM. Previous studies of concerning nonadiabatic dynamics on metal
surfaces introduce two distinct types of electron-nucleus forces in
addition to the adiabatic one \citep{bohnen1975friction,grumbach1994abinitio,hellsing1984electronic,baer2004realtime}.
The first type manifests as rapid fluctuations resembling a stochastic
process, while the second type involves dissipation, simulated as
a friction force with a friction constant determined by the electronic
structure. Since the molecular dynamics treats atomic nuclei as classical,
the Langevin dynamics approach \citep{tuckerman2010statistical,becca2017quantum}
can be used to handle nonadiabatic effects, utilizing fluctuating-dissipating
forces to impose the electronic temperature as the average value of
the atomic nuclei kinetic energy in a canonical ensemble. 

AIMD simulations for WDM need to consider both quantum Fermionic degeneracy
and strong Coulomb interactions \citep{karasiev2014innovations}.
The Kohn-Sham density functional theory (KS-DFT), which has proven
highly successful as an ab initio theory for elucidating the structure
of molecules and materials at zero temperature \citep{geerlings2003conceptual,martin2004electronic,hutter2012carparrinello,pribram-jones2015dfta,martin2016interacting,jensen2017introduction,vonlilienfeld2020exploring,perdew1983physical,kronik2012excitation,morales2010equation},
fulfills these requirements and has been extended to finite temperatures
and has become a widely employed method for theoretical studies of
WDM \citep{mermin1965thermal,pittalis2011exactconditions,pribram-jones2014thermal,groth2017abinitio,bonitz2020abinitio,moldabekov2022benchmarking}.
However, applying KS-DFT to WDM challenges calculating and storing
an increasing number of Kohn-Sham eigenstates as the temperature rises.
Consequently, for electronic temperatures exceeding $100,000\text{K}$,
other approaches, such as the extended KS method \citep{zhang2016extended,blanchet2022extended}
or \textquotedbl orbital-free DFT\textquotedbl , which include finite
temperature orbital free functionals \citep{feynman1949equations,perrot1979gradient,karasiev2012generalizedgradientapproximation,sjostrom2014fastand,karasiev2013nonempirical,luo2020towards}
and method development \citep{ligneres2005anintroduction,lambert2006veryhightemperature,ticknor2016transport,hu2017firstprinciples,white2018timedependent,white2019multicomponent,mi2023orbitalfree}
are preferred. Another emerging approach involves utilizing machine
learning to generate potential-energy surfaces and interatomic forces
based on KS-DFT and variational quantum Monte Carlo datasets. These
learned models can then be employed in molecular dynamics calculations
to predict material properties with reduced computational costs \citep{cheng2020evidence,vonlilienfeld2020retrospective,karasiev2021onthe,ellis2021accelerating,cheng2023thermodynamics}.

A linear scaling DFT procedure holds significant promise for investigating
WDM, whether applied directly in AIMD or for generating training data
sets for machine learning. This can be realized through stochastic
DFT, as demonstrated by various authors in recent works \citep{baer2013selfaveraging,cytter2018stochastic,cytter2019transition,white2020fastand,chen2021stochastic,baer2022stochastic,chen2023combining}.
In this paper, we elaborate on additional advancements in stochastic
plane-waves Kohn-Sham density functional theory, integrating it with
Langevin dynamics and introducing a novel approach for computing electronic
conductivity. We thoroughly assess and benchmark the method, showcasing
its practical application by conducting a detailed study of Hydrogen
at 30,000K.

\section{\label{sec:Stochastic-finite-temperature}Stochastic finite temperature
Kohn-Sham DFT }

\subsection{\label{subsec:Finite-temperature-Kohn-Sham-sch}Finite-temperature
Kohn-Sham scheme}

The combination of finite-temperature density functional theory \citep{hohenberg1964inhomogeneous,mermin1965thermal}
and the Kohn-Sham procedure (FT-KS-DFT) \citep{kohn1965selfconsistent}
greatly simplifies the formidable problem of treating interacting
electrons under the influence of a heat and electron bath. Under the
FT-KS-DFT formalism, we need only study a system of non-interacting
electrons in a one-body potential, which includes exchange-correlation
contributions. As a result of the exchange-correlation potential,
the non-interacting electron density is identical to that of the interacting
system, and the grand potential of the interacting electron system
can be expressed using the non-interacting grand potential together
with exchange-correlation free-energy corrections. As the finite temperature
diminishes towards zero, the FT-KS-DFT converges into the zero-temperature
KS-DFT, with corresponding exchange-correlation contributions, and
the free energy converges to zero-temperature ground-state energy
\citep{argaman2002thermodynamics}.

To study infinite systems, it is beneficial to impose periodic boundary
conditions within the simulation cell. The single electron wave functions
of the non-interacting system are expressed as a linear combination
of the plane wave basis $e^{i\boldsymbol{G}\cdot\boldsymbol{r}}$:
\begin{equation}
\psi\left(\boldsymbol{r}\right)=\sum_{\boldsymbol{G}}\tilde{c}_{\boldsymbol{G}}\frac{e^{i\boldsymbol{G}\cdot\boldsymbol{r}}}{\sqrt{\Omega}},\label{eq:psi(r)}
\end{equation}
where $\boldsymbol{G}=\frac{2\pi}{L}\left(m_{x},m_{y},m_{z}\right)$
is the simulation cell-commensurate wave vector and $m_{i}$ integers.
The wave vector parameter $G_{\text{cut}}$ determines the size of
the plane wave basis by requiring that $\left\Vert \boldsymbol{G}\right\Vert _{2}\le G_{\text{cut}}.$
This cutoff identifies a subspace of dimension $D=\left[\frac{3\pi}{4}\left(\frac{L}{2\pi}G_{\text{cut}}\right)^{3}\right]$
of the simulation cell's periodic functions, which is mapped by Eq.~\eqref{eq:psi(r)}
onto the complex vector space of D-tuples $\mathbb{C}^{D}=\left\{ \tilde{c}_{\boldsymbol{G}}\right\} _{\left\Vert \boldsymbol{G}\right\Vert _{2}\le G_{\text{cut}}}$.
The basis truncation error can be systematically mitigated by increasing
$G_{\text{cut}}$, or, equivalently the cutoff energy $E_{\text{cut}}=\frac{\hbar^{2}G_{\text{cut}}^{2}}{2m_{e}}$.
Variational treatment of the finite-temperature Kohn-Sham equations
within the subspace leads to a set of algebraic eigenvalue equations
\begin{equation}
\mathcal{H}\tilde{c}^{\left(j\right)}=\varepsilon_{j}\tilde{c}^{\left(j\right)},\quad j=1,2,\dots,D\label{eq:Eigen-value-equation}
\end{equation}
where $\mathcal{H}$ is the Kohn-Sham Hamiltonian (for more details
on the representation, and the operators see, for example, ref. \citep{martin2004electronic})
and $\varepsilon_{j}$ and $\tilde{c}^{\left(j\right)}$ are its (real)
eigenvalues and (complex) eigenvectors.

The electrons are in a grand canonical mixed state with temperature
parameter $\beta=\left(k_{B}T\right)^{-1}$, where $k_{B}$ is Boltzmann's
constant, and chemical potential $\mu$. The occupation of each single
particle energy level $\varepsilon$ is given by the Fermi-Dirac function
$p_{\beta\mu}\left(\varepsilon\right)\equiv\left(1+e^{\beta\left(\varepsilon-\mu\right)}\right)^{-1}$.
Correspondingly, the electron density can be expressed as a sum of
level densities, 
\begin{equation}
n\left(\boldsymbol{r}\right)=2\times\sum_{j}p_{\beta\mu}\left(\varepsilon_{j}\right)\left|\varphi_{j}\left(\boldsymbol{r}\right)\right|^{2},\label{eq:KS-density}
\end{equation}
where 
\begin{equation}
\varphi_{j}\left(\boldsymbol{r}\right)=\sum_{\boldsymbol{G}}\tilde{c}_{\boldsymbol{G}}^{\left(j\right)}\frac{e^{i\boldsymbol{G}\cdot\boldsymbol{r}}}{\sqrt{\Omega}}\label{eq:KS-orbital}
\end{equation}
are the (normalized, so that $\int_{\Omega}\left|\varphi_{j}\left(\boldsymbol{r}\right)\right|^{2}d\boldsymbol{r}=1$)
Kohn-Sham eigenstates in real-space. The grand potential of the electrons
is then given by \citep{weinert1992fractional} 
\begin{equation}
\mathit{\Phi}_{\beta\mu}\left[n\right]\equiv U\left[n\right]-\beta^{-1}S_{s}\left[n\right]-\mu\int n\left(\boldsymbol{r}\right)d\boldsymbol{r},\label{eq:def-KS-Phi}
\end{equation}
where 
\begin{align}
U\left[n\right] & \equiv E_{\text{orb}}-E_{H}\left[n\right]+\Phi_{\beta\mu,xc}\left[n\right]-\int v_{\beta\mu,xc}\left(\boldsymbol{r}\right)n\left(\boldsymbol{r}\right)d\boldsymbol{r}\label{eq:def-KS-U}
\end{align}
is the Kohn-Sham energy. In Eq.~\eqref{eq:def-KS-U}, 
\begin{equation}
E_{\text{orb}}\equiv2\times\sum_{j}p_{\beta\mu}^{j}\varepsilon_{j}\label{eq:def-KS-Eorb}
\end{equation}
is the orbital energy, $p_{\beta\mu}^{j}=p_{\beta\mu}\left(\varepsilon_{j}\right)$,
$E_{H}\left[n\right]$ is the Hartree energy, and $\Phi_{\beta\mu,xc}\left[n\right]$
is the $\beta\mu$-dependent approximate exchange-correlation free
energy functional of the density and $v_{\beta\mu,xc}\left(\boldsymbol{r}\right)=\frac{\delta\Phi_{\beta\mu,xc}}{\delta n\left(\boldsymbol{r}\right)}$.
While temperature-dependent exchange-correlation density functionals
have been developed recently \citep{karasiev2014accurate,karasiev2016importance,karasiev2018nonempirical,ramakrishna2020influence,hinz2020fullyconsistent,karasiev2022metagga},
in this paper we approximate $\Phi_{\beta\mu,xc}\left[n\right]$ by
the zero temperature local density approximation (LDA, \citep{kohn1965selfconsistent})
for the exchange correlation energy $E_{xc}^{LDA}\left[n\right]$
and correspondingly $v_{\beta\mu xc}\left(\boldsymbol{r}\right)$
is approximated as the zero temperature LDA exchange-correlation potential
$v_{xc}^{LDA}\left(\boldsymbol{r}\right)=\frac{\delta E_{xc}^{LDA}}{\delta n\left(\boldsymbol{r}\right)}$.
Finally, $S_{s}$ in Eq.~\eqref{eq:def-KS-Phi} is the entropy of
non-interacting electrons at density $n\left(\boldsymbol{r}\right)$
expressed as 
\begin{equation}
S_{s}=-2\times k_{B}\sum_{j}\left[p_{\beta\mu}^{j}\log p_{\beta\mu}^{j}+\bar{p}_{\beta\mu}^{j}\log\bar{p}_{\beta\mu}^{j}\right],\label{eq:def-NI-entropy}
\end{equation}
where, in brevity, $\bar{p}_{\beta\mu}^{j}\equiv1-p_{\beta\mu}^{j}$.
In this Kohn-Sham procedure we find the density $n\left(\boldsymbol{r}\right)$
that minimizes the grand potential $\mathit{\Phi}_{\beta\mu}$. The
number of electrons is $\tilde{N}_{e}\left(\mu\right)=\left(\frac{\partial\mathit{\Phi}_{\beta\mu}}{\partial\mu}\right)_{\beta}=2\times\sum_{j}p_{\beta\mu}^{j}$.

The $\xi$-component ($\xi=0,1,2$ indicates respectively $x,y,z$)
of the force on an atomic nucleus $A$ ( $A=0,\dots,N_{n}-1$, where
$N_{n}$ is the number of atomic nuclei in the simulation cell) is
equal to the corresponding derivative of the grand potential, $F^{i}=-\frac{\partial\mathit{\Phi}_{\beta\mu}}{\partial R_{i}}+F_{i}^{NN}$,
where $i\equiv\left(3A+\xi\right)$ is the force index, and $F_{i}^{NN}$
is the sum of forces exerted by all other atomic nuclei. This force
is an average force over all ground and excited electronic states
of all possible charge states of the system.

An alternative to working in the grand canonical ensemble, where $\mu$
is given, is to impose a fixed average number of electrons $N_{e}$
and then tune $\mu$ accordingly. Such an ensemble is more natural
for small, finite simulation cells. In this ensemble the chemical
potential becomes a function of the imposed value of $N_{e}$, denoted
$\tilde{\mu}\left(N_{e}\right)$, defined implicitly by solving the
equation
\begin{equation}
N_{e}=2\times\sum_{j}p_{\beta\tilde{\mu}\left(N_{e}\right)}^{j}.\label{eq:implcit-eq_miu-eigenvals}
\end{equation}
In this ensemble we find the density $n\left(\boldsymbol{r}\right)$
that minimizes the Helmholtz free energy $\mathcal{F}_{\beta N_{e}}=U-\beta^{-1}S_{s}$
and the force is its derivative, $F_{i}=-\frac{\partial\mathcal{F}_{\beta N_{e}}}{\partial R_{i}}+F_{i}^{NN}$.
Once again, this force is an average over all ground and excited electronic
states of all possible charge states of the system.

Regardless of the ensemble used, the electronic force component $i$
is obtained from the electron density and the corresponding derivative
of the electron-nucleus force potential: 
\[
F_{i}=-\int n\left(\boldsymbol{r}\right)\frac{\partial}{\partial R_{i}}v_{\text{eN}}\left(\boldsymbol{r}\right)d\boldsymbol{r}+F_{i}^{NN},
\]
and when non-local pseudopotentials are employed, i.e., $\hat{v}_{\text{eN}}=v_{\text{loc}}\left(\boldsymbol{r}\right)+\hat{v}_{nl}$
the following generalization needs be used (now in vector notation):
\begin{equation}
\boldsymbol{F}=-2\times\sum_{j}p_{\beta\mu}^{j}\left\langle \varphi_{j}\left|\boldsymbol{\nabla}\hat{v}_{\text{eN}}\right|\varphi_{j}\right\rangle +\boldsymbol{F}^{NN}.\label{eq:deterministic-force}
\end{equation}

\subsection{\label{subsec:The-stochastic-density}The stochastic density functional
approach}

Stochastic density functional theory \citep{baer2013selfaveraging}
is based on the concept of random wave functions 
\begin{equation}
\eta\left(\boldsymbol{r}\right)\equiv\sum_{\boldsymbol{G}}\tilde{\eta}^{\boldsymbol{G}}\frac{e^{i\boldsymbol{G}\cdot\boldsymbol{r}}}{\sqrt{\Omega}}\label{eq:def-eta(r)}
\end{equation}
in which the random coefficients $\tilde{\eta}^{\boldsymbol{G}}$
are given, in vector notation, by operating with the square-root Fermi-Dirac
operator $\sqrt{p_{\beta\mu}\left(\mathcal{H}\right)}$ on a random
vector: 
\begin{equation}
\tilde{\eta}\equiv\sqrt{p_{\beta\mu}\left(\mathcal{H}\right)}\tilde{\chi}\label{eq:def-eta(G)}
\end{equation}
where $\tilde{\chi}$ is a random vector with components $\tilde{\chi}^{\boldsymbol{G}}\equiv e^{i\theta_{\boldsymbol{G}}}$,
where $\theta_{\boldsymbol{G}}$ are independent random phases (between
$0$ and $2\pi$). It is straightforward to check that 
\begin{equation}
\text{E}\left[\tilde{\chi}^{\boldsymbol{G}'}\tilde{\chi}^{\boldsymbol{G}*}\right]=\delta_{\boldsymbol{G}'\boldsymbol{G}}.\label{eq:resolution-of-id}
\end{equation}
Here, the symbol $\text{E}\left[r\right]$ is the expected value of
a random variable $r$. The random variable $\tilde{\eta}^{\boldsymbol{G}'}\tilde{\eta}^{\boldsymbol{G}*}$
is an unbiased estimator of the KS density matrix in G-space, relying
on the following exact identity  $\left[p_{\beta\mu}\left(\mathcal{H}\right)\right]_{\boldsymbol{G}'\boldsymbol{G}}=\text{E}\left[\tilde{\eta}^{\boldsymbol{G}'}\tilde{\eta}^{\boldsymbol{G}*}\right]$.
This relation is proved by plugging Eq.~\eqref{eq:def-eta(G)} on
the right hand side and using Eq.~\eqref{eq:resolution-of-id}. Similarly,
sampling $\eta\left(\boldsymbol{r}\right)\eta\left(\boldsymbol{r}'\right)^{*}$,
where $\eta\left(\boldsymbol{r}\right)$ is defined in Eq.~\eqref{eq:def-eta(r)},
provides an estimate for the KS density matrix $\rho\left(\boldsymbol{r},\boldsymbol{r}'\right)=2\times\sum_{j}p_{\beta\mu}\left(\varepsilon_{j}\right)\varphi_{j}\left(\boldsymbol{r}\right)\varphi_{j}\left(\boldsymbol{r}'\right)^{*}$.
From this, $\left|\eta\left(\boldsymbol{r}\right)\right|^{2}$ is
an unbiased estimator for the electron density $n\left(\boldsymbol{r}\right)$,
relying on the exact identity:

\begin{equation}
n\left(\boldsymbol{r}\right)=2\times\text{E}\left[\left|\eta\left(\boldsymbol{r}\right)\right|^{2}\right].\label{eq:sKS-density}
\end{equation}
This expression is the essence of stochastic KS-DFT: it replaces the
\emph{calculation }of the electron density $n\left(\boldsymbol{r}\right)$
(Eq.~\eqref{eq:KS-density}), which requires the KS-DFT eigenstates
and eigenvalues (Eq.~\eqref{eq:Eigen-value-equation}) by a statistical
sampling of the random variable $\left|\eta\left(\boldsymbol{r}\right)\right|^{2}$.

The fact that the expected value of the absolute square of the random
variable $\eta\left(\boldsymbol{r}\right)$ gives the density, means
that we can now use sampling methods to obtain actual estimates of
the density. If we produce a sample of $I$ independent random vectors
$\tilde{\chi}_{i}$ ($i=1,\dots,I$) and from them, using Eqs.~(\ref{eq:def-eta(r)})-(\ref{eq:def-eta(G)})
obtain samples of $\eta_{i}\left(\boldsymbol{r}\right)$ then the
density can be estimated as an average 
\[
n\left(\boldsymbol{r}\right)=2\times\frac{1}{I}\sum_{i=1}^{I}\left[\left|\eta_{i}\left(\boldsymbol{r}\right)\right|^{2}\right].
\]
This sampling procedure is straightforward to parallelize using distributed
memory model, for example, the message-passing-interface library.
Observables, such as the forces on atomic nuclei, can be expressed
as stochastic traces as well (see subsection~\ref{sec:Stochastic-forces-and}).
From statistics, the fluctuations in the density or forces are proportional
to the inverse square root of the sample size. For the calculations
shown below we used a sample of $I=40$ stochastic orbitals (irrespective
of the system size).

\begin{figure*}
\includegraphics[width=0.47\textwidth]{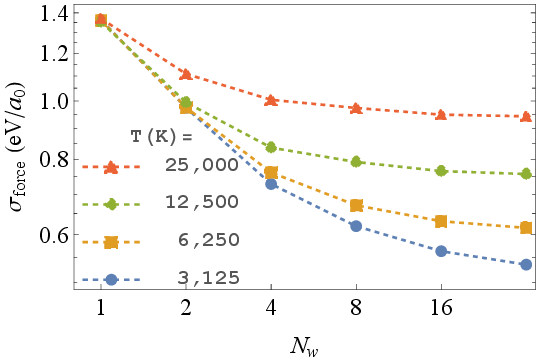} \includegraphics[width=0.47\textwidth]{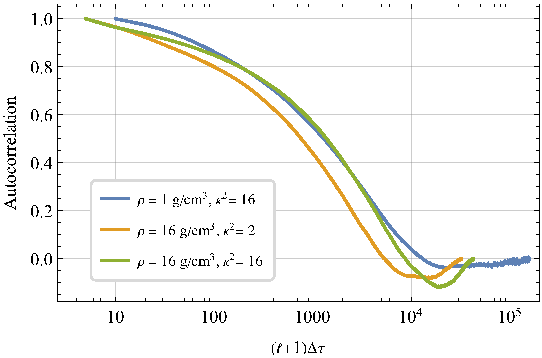}

\caption{\label{fig:sdft-force-stdev}\textbf{Left panel}: The sDFT force standard
deviation $\sigma_{f}\equiv\sqrt{\frac{1}{3N_{H}}\text{Tr}\Sigma_{\phi}^{2}}$
(per degree of freedom) for selected electronic temperatures as a
function of the number of windows $N_{w}$ in $\text{H}_{128}$ at
$\rho=1\,g/cm^{3}$. \textbf{Right panel}: The instantaneous position
autocorrelation $C_{R}\left(\ell\right)$ (see Eq.~\eqref{eq:sample-correlation},
averaged over all atoms) during a $T=30,000K$ Langevin trajectory
of $\text{H}_{256}$ in two densities with additional white noise
force (given in terms of $\kappa$, see Eq.~\eqref{eq:sDFT_and_white_noise-force-covar})
and a time step of $\Delta t=5\,\hbar E_{h}^{-1}$ for $\rho=16\,g/cm^{3}$
and $\Delta t=10\,\hbar E_{h}^{-1}$ for $\rho=1\,g/cm^{3}$.}
\end{figure*}

\subsection{\label{subsec:Chebyshev-expansion-methods}Chebyshev expansion methods}

\subsubsection{The essential Chebyshev expansion}

We now describe a recipe for performing calculations of the type shown
in Eq.~\eqref{eq:def-eta(G)}, i.e. operating with a function of
the Hamiltonian, namely $z\left(\mathcal{H}\right)$ on some given
a vector $\chi$: $\left|\zeta\right\rangle =z\left(\mathcal{H}\right)\left|\chi\right\rangle $.
For this, we use the Chebyshev expansion \citep{kosloff1988timedependent}
of length $N_{C}$, $\zeta^{\boldsymbol{G}}=\sum_{n=0}^{N_{C}-1}Z^{\left(n\right)}\chi_{n}^{\boldsymbol{G}}$,
which we write in ket-form as: 
\begin{equation}
\left|\zeta\right\rangle =\sum_{n=0}^{N_{C}-1}Z^{\left(n\right)}\left|\chi_{n}\right\rangle .\label{eq:ChebExpansion}
\end{equation}
The Chebyshev coefficients are defined by 
\[
Z^{\left(n\right)}=\frac{2}{N_{C}}e^{i\frac{\pi}{N_{C}}n}\tilde{z}^{\left(n\right)}
\]
where the series $\left\{ \tilde{z}^{\left(n\right)}\right\} _{n=0}^{N_{C}-1}$is
the discrete Fourier transform of $\left\{ z\left(\bar{E}+\Delta E\times\cos\left(\frac{l+\frac{1}{2}}{N_{C}}\pi\right)\right)\right\} _{l=0}^{N_{C}-1}$.
In the last expression, $\bar{E}=\left(E_{max}+E_{min}\right)/2$,
$\Delta E=\left(E_{max}-E_{min}\right)/2$ and $E_{min}$ ($E_{max}$)
is a lower (upper) bound estimate to the smallest (largest) eigenvalue
of $\mathcal{H}$.

The expansion length $N_{C}$ is chosen to be sufficiently large so
that the $\left|Z^{\left(n\right)}\right|$ are all smaller than a
threshold value $10^{-d}$, (typically $d=7$ or $8$) for $n>N_{C}$.
An estimate for the Chebyshev length is the following expression:
\begin{equation}
N_{C}\approx\frac{3d}{4}\times\beta\times\Delta E.\label{eq:NCheb}
\end{equation}

The Chebyshev vectors $\left|\chi_{n}\right\rangle $ in Eq.~\eqref{eq:ChebExpansion}
are defined as $\left|\chi_{n}\right\rangle \equiv T_{n}\left(\mathcal{H}_{s}\right)\left|\chi\right\rangle $,
where $T_{n}\left(x\right)$ is the $n^{th}$ Chebyshev polynomial
\citep{rivlin1990chebyshev} and $\mathcal{H}_{s}\equiv\frac{\mathcal{H}-\bar{E}}{\Delta E}$
is the shifted-scaled Hamiltonian, having all eigenvalues in the interval
$\left[-1,1\right]$. Based on a recurrence formula between any three
consecutive Chebyshev polynomials \citep{rivlin1990chebyshev}, the
Chebyshev vectors $\chi_{n}$ can be computed iteratively (hence only
three of them are needed at a given time): 
\begin{align*}
\left|\chi_{n}\right\rangle  & =2\mathcal{H}_{s}\left|\chi_{n-1}\right\rangle -\left|\chi_{n-2}\right\rangle ,\quad n\ge2,
\end{align*}
The first two vectors are given by: 
\[
\left|\chi_{0}\right\rangle =\left|\chi\right\rangle ,\qquad\left|\chi_{1}\right\rangle =\mathcal{H}_{s}\left|\chi_{0}\right\rangle ,
\]

\subsubsection{\label{subsec:Operating-with-several}Operating with several functions
of $\mathcal{H}$ on a given state $\left|\chi\right\rangle $ }

Each term in the Chebyshev expansion of Eq.~\eqref{eq:ChebExpansion}
is a product of a Chebyshev coefficient $Z^{n}$ and a Chebyshev vector
$\left|\chi_{n}\right\rangle $. The former depends on the function
$z\left(\mathcal{H}\right)$, while the latter does not. Suppose we
want to operate with several different functions $z_{m}\left(\mathcal{H}\right)$
($m=1,2,\dots,M$) on the \emph{same }vector $\chi$: 
\[
\left|\zeta_{m}\right\rangle =z_{m}\left(\mathcal{H}\right)\left|\chi\right\rangle .
\]
Chebyshev expansions can calculate these vectors 
\begin{equation}
\left|\zeta_{m}\right\rangle =\sum_{n=0}^{N_{C}-1}Z_{m}^{\left(n\right)}\left|\chi_{n}\right\rangle ,\label{eq:z_m-evaluation}
\end{equation}
where $Z_{m}^{\left(1\right)}$, $Z_{m}^{\left(2\right)}$, $\dots$
are the coefficients corresponding to the function $z_{m}\left(\varepsilon\right)$.
Most of the numerical effort goes into computing the vectors $\left|\chi_{n}\right\rangle $
and these are shared by all the different evaluations in Eq.~\eqref{eq:z_m-evaluation}.
Therefore, there is but a minute overhead in the effort to calculate
$M$ $\left|\zeta_{m}\right\rangle $s relative to just one $\left|\zeta\right\rangle $.

\subsubsection{\label{subsec:Energy-windows}Energy windows}

An example of using this approach is the \emph{Energy Windows }method
\citep{chen2019energywindow}. Here, we define $N_{w}$ chemical potentials
\[
\mu_{N_{w}}\equiv\mu\ge\mu_{N_{w}-1}\ge\dots\ge\mu_{1}
\]
and corresponding energy window projections 
\begin{align*}
z_{m}\left(\mathcal{H}\right) & =\sqrt{p_{\beta\mu_{m}}-p_{\beta\mu_{m-1}}},\qquad m=2,\dots,N_{w}\\
z_{1}\left(\mathcal{H}\right) & =\sqrt{p_{\beta\mu_{1}}}.
\end{align*}
Each of these functions projects a different energy range between
the chemical potentials. The sum of the square of these functions
yields the Fermi-Dirac projector 
\begin{align*}
p_{\beta\mu}\left(\mathcal{H}\right) & =z_{N_{w}}\left(\mathcal{H}\right)^{2}+z_{N_{w}-1}\left(\mathcal{H}\right)^{2}+\dots+z_{1}\left(\mathcal{H}\right)^{2}.
\end{align*}
Therefore, for any one-body operator $\mathcal{A}$, the KS expectation
value $\left\langle \mathcal{A}\right\rangle \equiv\text{Tr}\left[p_{\beta\mu}\left(\mathcal{H}\right)A\right]$
can be written as a sum of contributions from differing energy windows:
\[
\left\langle \mathcal{A}\right\rangle =\sum_{m=1}^{N_{w}}\text{Tr}\left[z_{m}\left(\mathcal{H}\right)Az_{m}\left(\mathcal{H}\right)\right].
\]
The equivalent stochastic expression is 
\[
\left\langle \mathcal{A}\right\rangle =\sum_{m=1}^{N_{w}}\text{E}\left[\left\langle \zeta_{m}\left|\mathcal{A}\right|\zeta_{m}\right\rangle \right].
\]
Depending on the observable $\mathcal{A}$, this procedure helps reduce
the fluctuations in estimating $\left\langle \mathcal{A}\right\rangle $
since $\zeta_{m}$ and $\zeta_{m'}$ span largely non-overlapping
energy windows.

In the left panel of Fig. \ref{fig:sdft-force-stdev}, we show the
standard deviation of the electronic force on the atomic nuclei (per
degree of freedom) as a function of the number of windows for selected
temperatures. It is seen that for low temperatures this standard deviation
is reduced by as much as a factor of $2$ as $N_{W}$ reaches $16$
or $32$. However, for the high temperature considered, the windows
are less efficient, reducing the standard deviation by, at most, a
factor of $1.4$.

\subsubsection{\label{subsec:Chebyshev-moments}Chebyshev moments }

A Chebyshev moment $M_{n}$ is the \emph{trace }of a Chebyshev polynomial
$T_{n}\left(\mathcal{H}_{s}\right)$. The overlap $\left\langle \chi\left|\chi_{n}\right.\right>$
is an unbiased estimator of $M_{n}$, based on the identity $\text{Tr}\left[T_{n}\left(\mathcal{H}_{s}\right)\right]=\text{E}\left[\left\langle \chi\left|T_{n}\left(\mathcal{H}_{s}\right)\right|\chi\right\rangle \right]$,
or: 
\[
M_{n}=\text{E}\left[\left\langle \chi\left|\chi_{n}\right.\right>\right].
\]
Knowledge of the moments allows us to compute the trace of any function
$z\left(\mathcal{H}\right)$ of the KS Hamiltonian $\mathcal{H}$
through the formula: 
\[
\text{Tr}\left[z\left(\mathcal{H}\right)\right]=\sum_{n=0}^{N_{C}-1}Z^{\left(n\right)}M_{n},
\]
where $Z^{\left(n\right)}$ are the coefficients for the Chebyshev
expansion of the function $z\left(\varepsilon\right)$.

Examples where moments are useful: 
\begin{enumerate}
\item When working in the canonical ensemble mentioned above, with a fixed
number of electrons $N_{e}$, the chemical potential $\mu$ is a function
of $N_{e}$ defined implicitly by Eq.~\eqref{eq:implcit-eq_miu-eigenvals},
depending on the KS eigenvalues $\varepsilon_{j}$. However, in sDFT,
we do not have access to $\varepsilon_{j}$. Hence, we use the Chebyshev
Moments to develop an alternative implicit equation for $\mu$: 
\begin{equation}
N_{e}=2\times\sum_{n=0}^{N_{C}-1}P_{\beta\mu\left(N_{e}\right)}^{\left(n\right)}M_{n},\label{eq:implcit-eq_miu-moments}
\end{equation}
where the $P_{\beta\mu}^{\left(n\right)}$s are the Chebyshev coefficients
corresponding to $p_{\beta\mu}\left(\varepsilon\right)$. The actual
determination of $\mu$ uses a numerical root-searching algorithm
(e.g., the bisection method) applied to Eq.~\eqref{eq:implcit-eq_miu-moments}.
The search for $\mu$ is a speedy step since the Chebyshev moments
are independent of $\mu$, so they are calculated only once and then
stored while calculating the Chebyshev coefficients $P_{\beta\mu}$
any value of $\mu$ only involves a single fast Fourier transform. 
\item The non-interacting electron entropy of Eq.~\eqref{eq:def-NI-entropy}
is estimated as 
\begin{equation}
S_{s}=2\times\sum_{n=0}^{N_{C}-1}S_{\beta\mu}^{\left(n\right)}M_{n},\label{eq:entropi-by-moments}
\end{equation}
where $S_{\beta\mu}^{\left(n\right)}$ are the Chebyshev coefficients
corresponding to the function $s_{\beta\mu}\left(\varepsilon\right)=-\left(p_{\beta\mu}\left(\varepsilon\right)\log p_{\beta\mu}\left(\varepsilon\right)+\bar{p}_{\beta\mu}\left(\varepsilon\right)\log\bar{p}_{\beta\mu}\left(\varepsilon\right)\right)$. 
\item The orbital energy $E_{\text{orb}}$ of Eq.~\eqref{eq:def-KS-Eorb}
is estimated as 
\[
E_{\text{orb}}=2\times\sum_{n=0}^{N_{C}-1}E_{\beta\mu}^{\left(n\right)}M_{n},
\]
where the $E_{\beta\mu}^{\left(n\right)}$ are the Chebyshev coefficients
corresponding to $e_{\beta\mu}\left(\varepsilon\right)=p_{\beta\mu}\left(\varepsilon\right)\varepsilon$. 
\end{enumerate}
\begin{figure}
\begin{centering}
\includegraphics[width=0.9\columnwidth]{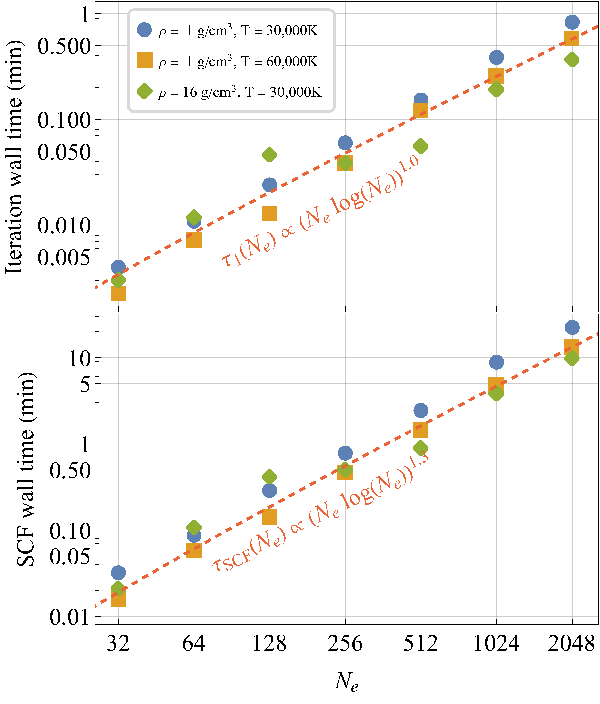} 
\par\end{centering}
\caption{\label{fig:Performance}The self-consistent-field (SCF) wall times
as a function of the number of electrons $N_{e}$ for Hydrogen at
the specified densities and temperatures. \textbf{Top panel}:\textbf{
}a single SCF iteration; \textbf{Bottom panel}: the entire SCF calculation
(stopped when the changes in energy per electron are below $10^{-5}E_{h}$).
The calculations used the cutoff energy of $E_{\text{cut}}=9E_{h}$,
40 stochastic orbitals and a \emph{single core per stochastic orbital
}on our Core-i7 cluster. }
\end{figure}

\subsection{\label{sec:Performance}Performance of the sDFT calculation}

Fig.~\ref{fig:Performance} shows the wall time self-consistent calculations
(averaged over many sDFT Langevin dynamics steps) as a function of
the number of electrons $N_{e}$ in the simulation cell for Hydrogen
at specified densities and temperatures. The computation time of a
single cycle of an SCF calculation scales linearly in $x=N_{e}\log N_{e}$.
Since the number of SCF cycles required to converge to a given criterion
grows mildly with system size, the overall scaling is $x^{1.3}$.
Calculations with higher temperatures and the same number of electrons
$N_{e}$ are faster since the Chebyshev expansions shorten in inverse
proportion to temperature (see Eq.~\eqref{eq:NCheb}). Higher density
calculations with the same number of atoms $N_{e}$ also require less
computation time because of the smaller simulation cell sizes.

Let us discuss the wall times and their dependence on scale the cutoff
energy $E_{\text{cut}}$. The plane wave basis size is determined
by the volume in $G$-space of the highest momentum vector $G_{max}=\sqrt{\frac{2m_{e}}{\hbar^{2}}E_{\text{cut}}}$
(where $m_{e}$ is the electron mass and $\hbar$ is Planck's constant)
and therefore proportional to $E_{\text{cut}}^{3/2}$. In addition,
the length of the Chebyshev expansion is proportional to $E_{\text{cut}}$
(see Eq.~\eqref{eq:NCheb}). Hence, overall, the wall time scales
steeply as $E_{\text{cut}}^{5/2}$, i.e., wall time increases by a
factor 32 every time the cutoff energy increases by a factor 4.

We have not yet developed a capability to expedite the calculation
speed for each stochastic orbital. We can achieve high factors if
we use a GPU on each node for this purpose.

\section{\label{sec:Stochastic-forces-and}Stochastic forces and Langevin
dynamics}

In the previous section, we discussed the WDM's electronic structure
at inverse temperature $\beta$. In this section we concentrate more
on the behavior of the atomic nuclei in WDM. At thermal equilibrium
their state is canonically distributed with a temperature identical
to that of the electrons. Here we discuss how we use the method of
Langevin dynamics to estimate the expected value of various observables
concerning atomic nuclei within the canonical ensemble.

\subsection{\label{subsec:Stochastic-forces-and}Regularization of the stochastic
forces}

In sDFT, the force on each atomic nucleus is a vector of a random
variable components \citep{baer2013selfaveraging}: 
\begin{equation}
\boldsymbol{f}=-2\times\text{\ensuremath{\left\langle \eta\left|\boldsymbol{\nabla}\hat{v}_{\text{eN}}\right|\eta\right\rangle }}+\boldsymbol{F}^{NN}\label{eq:random-force}
\end{equation}
where $\eta\left(\boldsymbol{r}\right)$ is defined in Eq.~\eqref{eq:def-eta(r)}
and $\boldsymbol{F}^{NN}$ is the force due to the other bare atomic
nuclei. Using the stochastic trace formula, the force $\boldsymbol{F}$
of Eq.~\eqref{eq:deterministic-force} is the expected value of the
random force: 
\[
\boldsymbol{F}=\text{E}\left[\boldsymbol{f}\right].
\]

The $3N_{n}\times3N_{n}$ symmetric positive-definite force covariance
matrix is $\left(\Sigma_{f}^{2}\right)_{ii'}=\text{E}\left[f_{i}f_{i'}\right]-F_{i}F_{i'}$,
or, in matrix notation 
\begin{equation}
\Sigma_{f}^{2}\equiv\text{E}\left[\boldsymbol{f}\boldsymbol{f}^{T}\right]-\boldsymbol{F}\boldsymbol{F}^{T}.\label{eq:sDFT-force-covar}
\end{equation}
To ease the handling of stochastic forces, we add independent white
noise $\boldsymbol{\zeta}$ ($\text{E}\left[\boldsymbol{\zeta}\right]=0$):
\begin{equation}
\boldsymbol{\boldsymbol{\varphi}}=\boldsymbol{f}+\boldsymbol{\zeta}.\label{eq:random-force-with-white-noise}
\end{equation}
The covariance matrix $\text{E}\left[\boldsymbol{\zeta}\boldsymbol{\zeta}^{T}\right]$
of this additional white noise is specially constructed to allow the
covariance $\Sigma_{\boldsymbol{\varphi}}^{2}$ of the total force
to be uniform, i.e. a \emph{multiple of the unit matrix}: 
\begin{equation}
\Sigma_{\boldsymbol{\varphi}}^{2}\equiv\text{E}\left[\boldsymbol{\varphi}\boldsymbol{\varphi}^{T}\right]-\boldsymbol{F}\boldsymbol{F}^{T}=\sigma^{2}\text{I}=\kappa^{2}\sigma_{f}^{2}\text{I},\label{eq:sDFT_and_white_noise-force-covar}
\end{equation}
where $\sigma^{2}$ is larger ($\kappa\ge1)$ than the largest eigenvalue
$\sigma_{f}^{2}$ of $\Sigma_{f}^{2}$. The white noise force $\zeta$
is thus sampled, using the Metropolis-Hastings algorithm, to have
a Gaussian distribution with the positive-definite covariance matrix
$\Sigma_{\zeta}^{2}\equiv\sigma^{2}I-\Sigma_{f}^{2}$. The procedure
obviously requires an estimate of the sDFT force covariance $\Sigma_{f}^{2}$
and we use a sample of $N_{s}$ force vectors to estimate it (with
$N_{s}=50$, seemingly quite sufficient). The covariance estimate
is done once every $N_{j}$ MD steps (we took $N_{j}=150$). A force
having such a uniform covariance matrix enables using the same friction
coefficient for all degrees of freedom, and therefore simplifies the
temperature control in the Langevin dynamics calculation.

\subsection{The stochastic Langevin equations of motion}

The stochastic force $\boldsymbol{\boldsymbol{\varphi}}\left(\boldsymbol{R}\right)$
for a given atomic nuclei configuration is now used to perform Langevin
molecular dynamics from which we obtain configuration and momentum
samples that are canonically distributed. From these samples we can
compute the thermodynamic properties of the system. The dynamics involves
solving the Langevin equation of motion 
\begin{align*}
\dot{\boldsymbol{P}}\left(t\right) & =\boldsymbol{\boldsymbol{\varphi}}\left(\boldsymbol{R}\left(t\right)\right)-\gamma\boldsymbol{P}\left(t\right),\\
\dot{\boldsymbol{R}}\left(t\right) & =M^{-1}\boldsymbol{P}\left(t\right),
\end{align*}
where $M^{-1}$ is a diagonal matrix of the inverse nuclei mass, and
$\gamma$ is the diagonal matrix of friction coefficients. We use
a time-discretized solver \citep{luo2014abinitio} for the stochastic
differential equation, from which we obtain a discretized trajectory
of $N_{T}$ atomic configurations $\boldsymbol{R}^{\left(n\right)}=\boldsymbol{R}\left(n\Delta\tau\right)$
($n=1,\dots,N_{T}$) and their momenta $\boldsymbol{P}^{\left(n\right)}=\boldsymbol{P}\left(\left(n-\frac{1}{2}\right)\Delta\tau\right)$,
where $\Delta\tau$ is the time step. The phase-space trajectory is
built from the following evolution steps 
\begin{align*}
\boldsymbol{P}^{\left(n+1\right)} & =e^{-\gamma\Delta\tau}\boldsymbol{P}^{\left(n\right)}+\left(\frac{1-e^{-\gamma\Delta\tau}}{\gamma}\right)\boldsymbol{\boldsymbol{\varphi}}\left(\boldsymbol{R}^{\left(n\right)}\right),\\
\boldsymbol{R}^{\left(n+1\right)} & =\boldsymbol{R}^{\left(n\right)}+M^{-1}\boldsymbol{P}^{\left(n+1\right)}\Delta\tau,
\end{align*}
and in the limit $\Delta\tau\to0$ a Langevin trajectory is obtained.
Here, the (diagonal) friction matrix $\gamma$ is determined from
the fluctuation-dissipation relation, which is given by 
\begin{equation}
\sigma^{2}=\frac{\gamma\Delta\tau/2}{\tanh\left(\gamma\Delta\tau/2\right)}\times\frac{2\gamma M}{\beta}.\label{eq:fluctuationDisspation}
\end{equation}

\subsection{\label{subsec:Statistical-sampling}Statistical sampling}

For sufficiently small $\Delta\tau$ each of the trajectory configurations
$\boldsymbol{R}^{\left(n\right)}$ is equivalent to a sample taken
from the Boltzmann distribution $p_{\beta}^{B}\left(\boldsymbol{R}\right)\propto e^{-\beta V_{BO}\left(\boldsymbol{R}\right)}$
where $V_{BO}\equiv\mathit{\Phi}_{\beta\mu}+E_{NN}$ ($V_{BO}=\mathcal{F}_{\beta N_{e}}+E_{NN}$)
is the \emph{electronic }grand-canonical (canonical) potential (see
Eq.~\eqref{eq:def-KS-Phi}). The momentum $\boldsymbol{P}^{\left(n\right)}$
is equivalent to a sample from the Maxwell-Boltzmann probability distribution
function $p_{R\beta}^{MB}\left(\boldsymbol{P}\right)\propto e^{-\beta\sum_{i=1}^{N_{N}}\frac{\boldsymbol{P}_{i}^{2}}{2M_{i}}}$.

The estimate for the thermal average $\left\langle O\right\rangle _{\beta}\equiv\iint O\left(\boldsymbol{R},\boldsymbol{P}\right)p_{\beta}^{B}\left(\boldsymbol{R}\right)d\boldsymbol{R}p_{\beta}^{MB}\left(\boldsymbol{P}\right)d\boldsymbol{P}$
of a given observable $O\left(\boldsymbol{R},\boldsymbol{P}\right)$
is simply the sample mean $\bar{O}\equiv\frac{1}{N_{T}}\sum_{n=1}^{N_{T}}O^{\left(n\right)}$
over the sequence $O^{\left(n\right)}\equiv O\left(\boldsymbol{R}^{\left(n\right)},\boldsymbol{P}^{\left(n\right)}\right)$.
The sample variance $\overline{\Delta O^{2}}=\frac{1}{N_{T}}\sum_{n=1}^{N_{T}}\Delta O^{\left(n\right)}$,
where $\Delta O^{\left(n\right)}=O^{\left(n\right)}-\bar{O}$, allows
us to determine a confidence interval for the thermal average. For
example, the 70\% confidence interval is $\left[\bar{O}-\delta O,\bar{O}+\delta O\right]$
where $\delta O\equiv\sqrt{\frac{1}{N_{\text{ind}}}\overline{\Delta O^{2}}}$and
$N_{\text{ind}}$ is the number of statistically independent samples
in the sequence $O^{\left(n\right)}$. If the values $O^{\left(1\right)}$,
$O^{\left(2\right)}$, $\dots$ were uncorrelated then $N_{\text{ind}}$
would be just the sample size $N_{T}$. However, because the configurations
$\boldsymbol{R}^{\left(n\right)}$ are part of a \emph{molecular dynamics
trajectory}, $O^{\left(n+1\right)}$ is \emph{correlated }with $O^{\left(n\right)}$,
and $O^{\left(n+2\right)}$ is correlated with $O^{\left(n+1\right)}$,
etc. and therefore $N_{ind}<N_{T}$. It is common to quantify the
strength of this correlation using \emph{the auto-correlation function}
for $O$, defined by $C_{O}\left(\ell\Delta\tau\right)\equiv\frac{\left\langle \Delta O^{\left(n\right)}\Delta O^{\left(n+\ell\right)}\right\rangle }{\left\langle \Delta O^{\left(n\right)2}\right\rangle }$
(the expression on the right-hand side is independent of $n$). In
a given sample trajectory, the auto-correlation function is estimated
by 
\begin{equation}
C_{O}\left(\ell\Delta\tau\right)\approx\frac{\sum_{n}\Delta O^{\left(n\right)}\Delta O^{\left(n+\ell\right)}}{\sum_{n}\Delta O^{\left(n\right)2}}.\label{eq:sample-correlation}
\end{equation}
It starts with the value $C_{O}\left(0\right)=1$ (full correlation)
and then decays steadily as step separation $\ell$ grows until hitting
a regime of small random fluctuations. The decay is characterized
by a correlation time $\tau_{O}$, for which $C_{O}\left(\tau_{O}\right)=e^{-1}$.
We also define the \emph{correlation length} $\ell_{O}=\frac{\tau_{O}}{\Delta\tau}$.
We view $\ell_{O}$ consecutive samples as ``correlated'' while
later samples are considered \emph{uncorrelated}.\emph{ }The number
of \emph{effectively independent }samples is thus estimated as: $N_{\text{ind}}\approx N_{T}/\ell_{O}$.
In the right panel of Fig.~\ref{fig:sdft-force-stdev} we show the
Langevin dynamics position autocorrelation function $C_{R}\left(\ell\Delta\tau\right)$
of hydrogen at $T=30,000K$, for two densities $\rho$ and two white
noise parameters $\kappa^{2}$ (see Eq.~\eqref{eq:sDFT_and_white_noise-force-covar}).
The correlation times $\tau_{R}$ are weakly dependent on the density
but grow significantly with $\kappa$. Hence, we strive for small
values of $\kappa^{2}>1$.

\subsection{Computational demonstration of $\mu VT$ - $N_{e}VT$ ensemble equivalence }

\begin{figure}
\begin{centering}
\includegraphics[width=0.5\columnwidth]{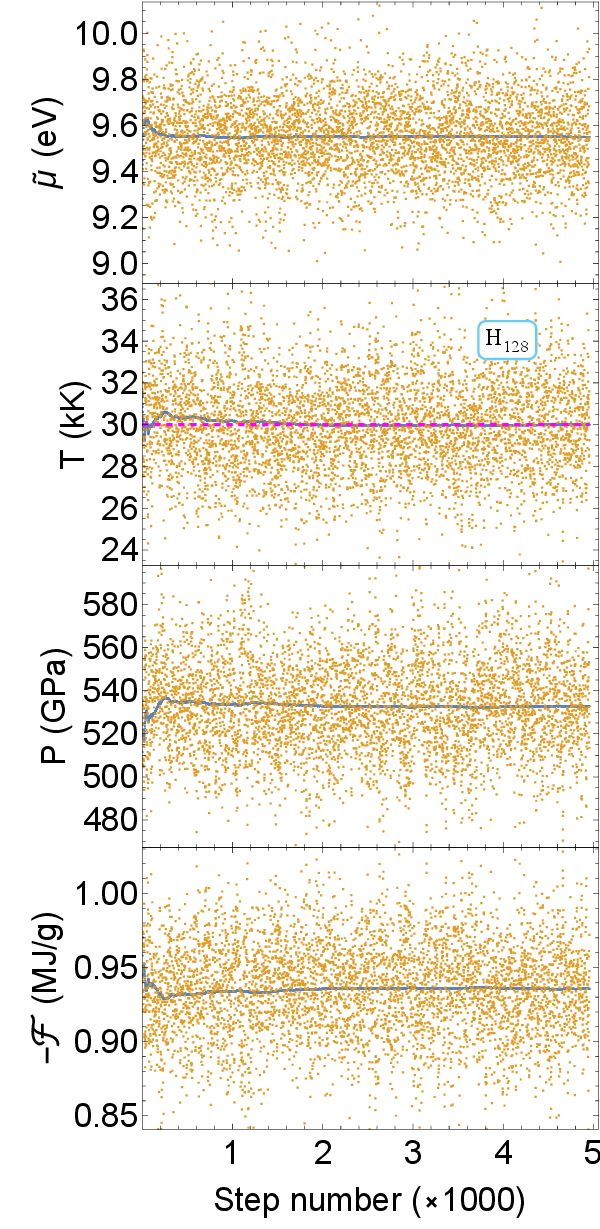}\includegraphics[width=0.5\columnwidth]{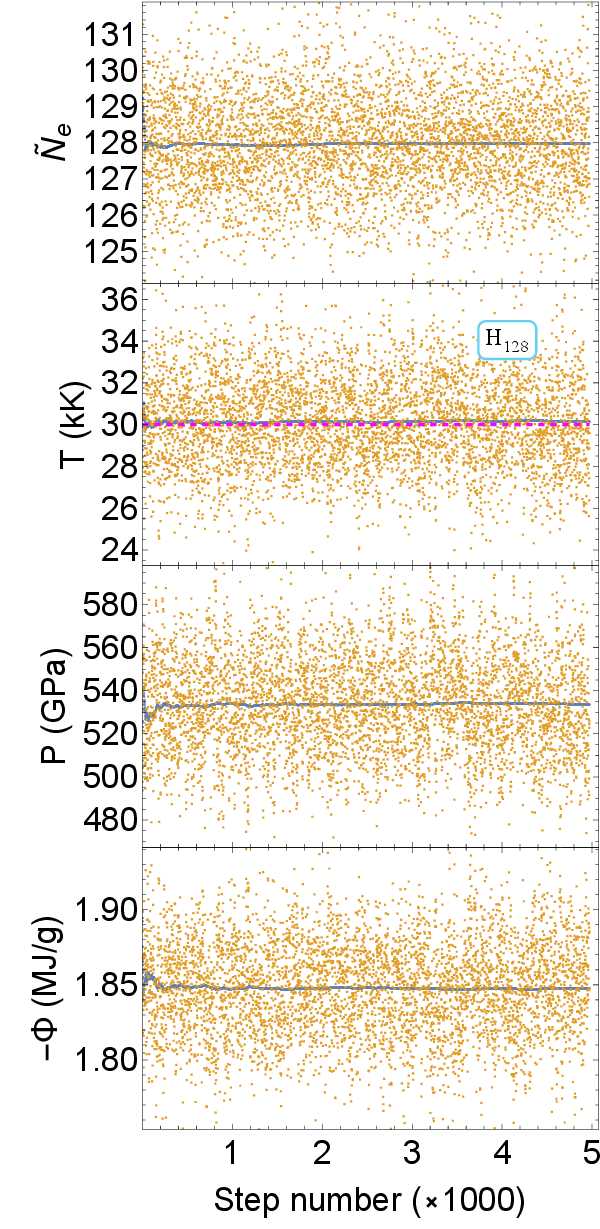} 
\par\end{centering}
\caption{\label{fig:TrajectoryH128}The instantaneous values (brown) and running
averages (blue) of observables in two Langevin molecular dynamics
trajectories of $\text{H}_{128}$ (at mass density of $\rho=1\text{g}\times\text{cm}^{-3}$
and target temperature of $30,000\text{K}$): on the \textbf{left
panel}, the NVT-like trajectory, where the number of electrons is
fixed ($N_{e}=128$) by tuning the chemical potential $\tilde{\mu}^{\left(n\right)}$
at each time step (see Eq.~\eqref{eq:implcit-eq_miu-eigenvals}).
On the \textbf{right panel}, the $\mu VT$ trajectory, where the chemical
potential is fixed ($\mu=9.56\,\text{eV}$) while the number of electrons
$N_{e}^{\left(n\right)}$ fluctuates. Details of the Langevin dynamics:
the white noise fluctuation is $\sigma=4\sigma_{f}$, i.e. $\kappa=4$
(see Eq.~\eqref{eq:sDFT_and_white_noise-force-covar}), the time
step is $\Delta\tau=10\,\text{atu}$, and the sDFT force covariance
(see Eq.~\eqref{eq:sDFT-force-covar}) is calculated using $50$
independent samples once every 150 dynamical time steps.}
\end{figure}

Fig.~\ref{fig:TrajectoryH128} displays time-dependent values of
selected observables in two Langevin dynamics trajectories of hydrogen
at mass density $\rho=1\,g\times cm^{-3}$ at $T=30,000K$. The two
trajectories are calculated in different electronic ensembles: the
left panel of the plot shows the results of an $N_{e}VT$-like ensemble,
where we impose a constant electron number $N_{e}=128$ at each time
step by tuning the electronic chemical potential $\tilde{\mu}\left(N_{e};\boldsymbol{R}^{\left(n\right)}\right)$
in the Fermi-Dirac function at each time step (see Eq.~\eqref{eq:implcit-eq_miu-eigenvals}).
This chemical potential fluctuates in time, as do the positions and
momenta of the atomic nuclei. The right panel shows the results of
an $\mu VT$ ensemble, where the electronic chemical potential is
set to a constant value of $\mu=9.56\,eV$. Now the number of electrons
fluctuate but \emph{on average }it is 128. The observables are the
kinetic energy per atomic degree of freedom $T^{\left(n\right)}$
(divided by $k_{B}$ and given in kilo-Kelvin), the pressure $P^{\left(n\right)}$,
the chemical potential $\tilde{\mu}^{\left(n\right)}$, the Helmholtz
energy $\mathcal{\tilde{F}}_{\beta N_{e}}^{\left(n\right)},$ in the
left panel, the electron number $\tilde{N}_{e}^{\left(n\right)}$
and the Grand potential $\tilde{\mathit{\Phi}}_{\beta\mu}^{\left(n\right)}$
in the right panel. Upon studying the numerical results in Fig.~\ref{fig:TrajectoryH128},
it is obvious that the average over the fluctuating chemical potential
on the left panel $\left\langle \tilde{\mu}\right\rangle _{N_{e}VT}=\frac{1}{N_{T}}\sum_{n=1}^{N_{T}}\tilde{\mu}\left(N_{e};\boldsymbol{R}^{\left(n\right)}\right)$
is very similar to the constant chemical potential $\mu$ imposed
in the $\mu VT$ ensemble on the right panel. Similarly, the average
over the fluctuating number of electrons on the right panel $\left\langle \tilde{N}_{e}\right\rangle _{\mu VT}=\frac{1}{N_{T}}\sum_{n=1}^{N_{T}}\tilde{N}_{e}\left(\mu;\boldsymbol{R}^{\left(n\right)}\right)$
is very close to the imposed number of electrons $N_{e}$ used in
the $N_{e}VT$ ensemble on the left panel. These results can be summarized
in the following relation: 
\[
\mu=\left\langle \tilde{\mu}\right\rangle _{N_{e}VT}\Leftrightarrow N_{e}=\left\langle \tilde{N}_{e}\right\rangle _{\mu VT},
\]
showing that in our finite-sized system, the two ensembles $\mu VT$
and $N_{e}VT$ are already equivalent, which is characteristic of
the thermodynamic limit. All calculations shown in the next section
were performed in the $N_{e}VT$ ensemble. 

\section{Kubo-Greenwood conductivity}

In this section, we consider the stochastic calculation of the Kubo-Greenwood
conductivity \citep{kubo1957statisticalmechanical,greenwood1958theboltzmann}.
In the context of WDM, these calculations were addressed in refs.~\citep{holst2011electronic,calderin2017kubogreenwood}
but they become demanding as the system size and temperature increase.
Hence, a stochastic calculation may be preferable for such systems
as discussed in Ref.~\citep{cytter2019transition}. Here, we provide
an improved approach including the DC conductivity with considerably
lower statistical errors. We also provide a detailed description of
the theory, the derivation, and how the calculations were made.

\begin{figure*}
\begin{centering}
\includegraphics[height=0.31\textheight]{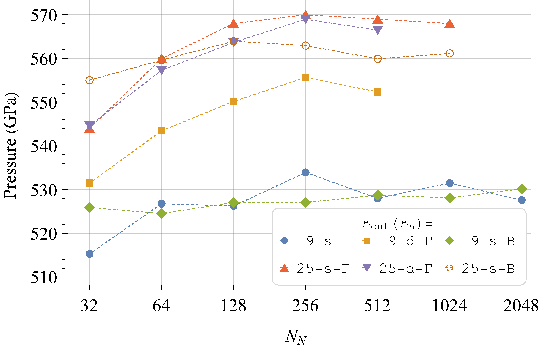}\includegraphics[height=0.31\textheight]{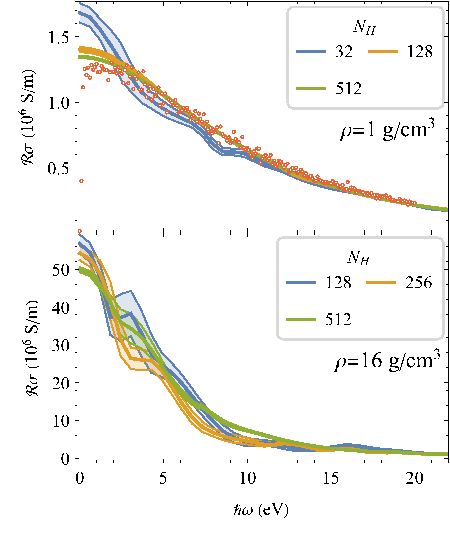} 
\par\end{centering}
\caption{\label{fig:Size-Dependence}Pressure and conductivity averaged over
a Langevin trajectory. \textbf{Left panel}: Size-dependence of pressure
at density $1\,\text{g}/\text{cm}^{3}$. Each point on the graph is
an average over the pressure estimates along the xDFT (x=''s'' or
``d'') trajectory, performed using $9$ or $25\,\text{E}_{\text{h}}$
cutoff energy at the $\Gamma$ or B (Baldereschi) k-point, as indicated
in the legend. The sDFT calculations for each point on the Langevin
trajectory used 40 stochastic orbitals at the LDA/NCPPs level. The
dDFT calculations were performed by the VASP program \citep{kresse1993abinitio,kresse1994abinitio,kresse1996efficient,kresse1996efficiency,kresse1999fromultrasoft},
at the LDA/PAW level. The trajectory time-step was $\Delta\tau=5\text{atu}$,
and white noise parameter $\kappa^{2}=2$. \textbf{Right panels}:
Size-dependence of the conductivity where the top (bottom) panel shows
results for $\rho=1\left(16\right)\text{g}/\text{cm}^{3}$. Each conductivity
curve is an average over the conductivity curves calculated for $N_{s}=20$
configurations of the atomic nuclei (snapshots) taken every 1000 atu
along the sDFT/Langevin trajectory. The error bars are $\pm s/\sqrt{N_{s}}$,
where $s$ is the standard deviation. One conductivity calculation
produces an entire conductivity curve ($\sigma\left(\omega\right)$)
based on 120 stochastic orbitals, performed at the Baldereschi k-point,
with a $15E_{h}$ cutoff energy. Dark empty circles appearing in the
top-right panel are conductivity calculations for $\text{H}_{128}$
on the identical configurations using the deterministic conductivity
method of Ref.~\citep{holst2011electronic} using VASP.}
\end{figure*}

Kubo's analysis \citep{kubo1957statisticalmechanical} starts with
expressing the complex conductivity 
\begin{equation}
\sigma_{\xi\xi'}\left(\omega\right)=\int_{0}^{\infty}\phi_{\xi\xi'}\left(t\right)e^{-i\omega t}dt,\label{eq:conductivity-from-response}
\end{equation}
as the Fourier transform of the dipole-current-density response function:
\[
\phi_{\xi\xi'}\left(t\right)=\frac{1}{i\hbar}\text{tr }\left(\rho\left[\sum_{n}e\mathcal{R}_{n\xi},\sum_{n'}\frac{e\mathcal{V}_{n'\xi'}\left(t\right)}{\Omega}\right]\right),
\]
where $e$ is the electron charge, $\text{tr}$ is a many-body trace,
$\rho$ is the equilibrium (many-body) density matrix, $\mathcal{R}_{n\xi}$
is the position in Cartesian direction $\xi$ ($\text{\ensuremath{\xi=x,y,z}}$)
of electron $n$, and $\mathcal{V}_{n\xi}\equiv\frac{\hbar}{m_{e}}\left(\frac{1}{i}\frac{\partial}{\partial\mathcal{R}_{n\xi}}-k_{\xi}\right)$
is the corresponding velocity (where $k_{\xi}=\frac{\pi}{2L_{\xi}}$
is the Baldereschi k-point). For non-interacting electrons, the response
function reduces to a single electron expression (see Appendix~\eqref{sec:proof-eq-phi}):
\begin{equation}
\phi_{\xi\xi'}\left(t\right)=\frac{4e^{2}}{\hbar\Omega}\Im\text{Tr }\left(p_{\mu\beta}\left(\mathcal{H}\right)\mathcal{R}_{\xi}\mathcal{V}_{\xi'}\left(t\right)\right),\label{eq:phi_xi_xi'}
\end{equation}
where $\text{Tr}$ is a single particle operator trace, $\mathcal{R}_{\xi}$
and $\mathcal{V}_{\xi'}$ are single electron position and velocity
operators respectively, $p_{\mu\beta}\left(\mathcal{H}\right)$ is
the Fermi-Dirac distribution, and $\mathcal{H}$ is the single particle
Hamiltonian (at the Baldereschi k-point), which we take from KS-DFT.
We have also included a factor 2 due to spin-degeneracy. The use of
such a static KS Hamiltonian, as opposed to the TDDFT description,
is the central approximation of the Kubo-Greenwood theory.

For $\omega\ne0$, we multiply and divide by $-i\omega$ the integral
of Eq.~\eqref{eq:conductivity-from-response}, use the identity $-i\omega e^{-i\omega t}=\frac{d}{dt}e^{-i\omega t}$
and then integrate by parts, obtaining 
\begin{equation}
\sigma_{\xi\xi'}\left(\omega\right)=\frac{1}{i\omega}\left(-\phi_{\xi\xi'}\left(0\right)+\int_{0}^{\infty}\dot{\phi}_{\xi\xi'}\left(t\right)e^{-i\omega t}dt\right),\label{eq:sigma-AC}
\end{equation}
which involves the velocity-velocity response function: 
\begin{align*}
\dot{\phi}_{\xi\xi'}\left(t\right) & =-\frac{4e^{2}}{\hbar\Omega}\Im\text{Tr}\left(p_{\mu\beta}\left(H\right)\mathcal{V}_{\xi}\mathcal{V}_{\xi'}\left(t\right)\right).
\end{align*}
For evaluating the trace, we use the stochastic trace formula: 
\begin{equation}
\text{Tr}\left(p\mathcal{V}_{\xi}\mathcal{V}_{\xi'}\left(t\right)\right)=\mathbb{E}\left[\left\langle \zeta_{\xi t}\left|\mathcal{V}_{\xi'}\right|\eta_{t}\right\rangle \right],\label{eq:stochstic-corr}
\end{equation}
where, for brevity, $p=p_{\beta\mu}\left(\mathcal{H}\right)$ and
\[
\left|\eta_{t}\right\rangle \equiv e^{-i\mathcal{H}t/\hbar}\sqrt{p}\left|\chi\right\rangle ,\qquad\left|\zeta_{\xi t}\right\rangle \equiv e^{-i\mathcal{H}t/\hbar}\mathcal{V}_{\xi}\left|\eta\right\rangle ,
\]
and $\left|\chi\right\rangle $ is a stochastic state. To use Eq.~\eqref{eq:stochstic-corr}
we generate a sample of $N_{s}$ stochastic vectors $\chi$, and for
each, we obtain a specific value of $\left\langle \zeta_{\xi t}\left|\mathcal{V}_{\xi'}\right|\eta_{t}\right\rangle $.
Averaging these values gives an estimate of the trace in the response
function with a statistical error proportional to $1/\sqrt{N_{s}}$.

The trace operations provide correlation functions, which we denote
$\dot{\phi}\left(t\right)$, (whether $\dot{\phi}_{\xi\xi'}\left(t\right)$
described above or $\dot{\psi}_{\xi\xi'}\left(t\right)$ described
below). To use it for obtaining the conductivity as a function of
$\omega$, we first select a desired spectral energy resolution $\hbar\nu$,
which defines a frequency grid $\omega_{g}=g\times\nu$, where $g=0,1,\dots,N_{\omega}$,
and then perform the Fourier integral of Eq\@.~\eqref{eq:sigma-AC}
for these frequencies. Given the resolution, the integral of the correlation
function $\dot{\phi}\left(t\right)$ is augmented by a Gaussian window,
discretized, and summed utilizing the fast-Fourier algorithm 
\begin{align}
\int_{0}^{\infty}\dot{\phi}\left(t\right)e^{-i\omega_{g}t}dt & \to\int_{0}^{\infty}e^{-\frac{\nu^{2}t^{2}}{2}}\dot{\phi}\left(t\right)e^{-i\omega_{g}t}dt\label{eq:discrete-integral}\\
 & \to\frac{\tau_{f}}{N_{\omega}}\sum_{g'=0}^{N_{\omega}}w_{g'}e^{-\frac{\nu^{2}\tau_{g'}^{2}}{2}}\dot{\phi}\left(\tau_{g'}\right)e^{-i\omega_{g}\tau_{g'}}\nonumber 
\end{align}
on an equally spaced time-grid $\tau_{g}=g\times\frac{\tau_{f}}{N_{\omega}}$,
extending from zero to $\tau_{f}=7/\nu$. The number of time and frequency
grid points is taken as $N_{\omega}=q_{\text{fac}}\times\frac{E_{\text{cut}}}{\hbar\nu}$,
where the quality factor $q_{\text{fac}}>1$ determines the precision
of the time integration. We also inserted integration weights, the
simplest of which is the trapeze rule: $w_{k}=\left(1-\delta_{k0}/2\right)$.
These weights are essential as the integral is a half-Fourier transform,
which means that the integrand does not decay smoothly to zero at
the time-grid boundaries. We experimented with various choices of
$\nu$ and $q_{fac}$, finding that a resolution of $\hbar\nu=0.025\text{E}_{\text{h}}$
and a quality factor in the range of $3$ to $5$ yield meaningful
and stable results. Dividing these values by $i\omega_{g}$ ($g>0$),
we obtain the AC conductivity $\sigma_{\xi\xi'}\left(\omega_{g}\right)$.

Since evaluating the conductivity in Eq\@.~\eqref{eq:sigma-AC}
requires division of the Fourier integral by $\omega$, the statistical
fluctuations are amplified as $\omega\to0$. The procedure is undefined
for $\omega=0$, the DC limit. In this case, we could use the analytical
limit of Eq\@.~\eqref{eq:sigma-AC}, 
\begin{equation}
\Re\sigma_{\xi\xi'}\left(0\right)=\int_{0}^{\infty}\dot{\phi}_{\xi\xi'}\left(t\right)tdt,\label{eq:sigma-DC-OPT 1}
\end{equation}
which does not divide by zero. We can use the same integration procedure
outlined above for the integral. But experience shows that the stochastic
error in Eq.~\eqref{eq:sigma-DC-OPT 1}, although finite, it is not
small and requires extensive sampling. For the important case of $\xi=\xi'$
it is possible to show (see Appendix \eqref{sec:proof-eq-psi}) that
\begin{equation}
\Re\sigma_{\xi\xi}\left(0\right)=\int_{0}^{\infty}\dot{\psi}_{\xi\xi}\left(t\right)dt,\label{eq:sigma-DC-OPT 2}
\end{equation}
where (see Eq.~\eqref{eq:final-proof}): 
\[
\dot{\psi}_{\xi\xi}\left(t\right)=-\frac{2e^{2}}{\Omega}\,\Re\,\text{Tr}\left(p_{\mu\beta}^{\prime}\left(H\right)\mathcal{V}_{\xi}\mathcal{V}_{\xi}\left(t\right)\right),
\]
and $p_{\mu\beta}^{\prime}\left(\varepsilon\right)=-\beta p_{\mu\beta}\left(\varepsilon\right)\left(1-p_{\mu\beta}\left(\varepsilon\right)\right)$
is the derivative of the Fermi-Dirac function. The stochastic evaluation
of the correlation function $\text{Tr}\left(p_{\mu\beta}^{\prime}\left(H\right)\mathcal{V}_{\xi}\mathcal{V}_{\xi}\left(t\right)\right)$
follows the same procedure as $\text{Tr}\left(p_{\mu\beta}\left(H\right)\mathcal{V}_{\xi}\mathcal{V}_{\xi}\left(t\right)\right)$
except that this time we take $p=p_{\beta\mu}^{\prime}\left(\mathcal{H}\right)$.
Then, the same time-integration scheme outlined above for the Fourier
integral of $\dot{\phi}_{\xi\xi}$ can be used to evaluate the integral
of Eq.~\eqref{eq:sigma-DC-OPT 2}.

The problem of high fluctuations in the low-frequency part of the
AC conductivity is exacerbated due to the need to take smaller $\nu$
for converging the AC results to the DC limit: the smaller $\nu$,
the larger the fluctuations. Thus, we developed an interpolation procedure
for the low-frequency spectrum relying on the steadiness of the DC
conductivity. This procedure is described in Appendix~\ref{sec:Stabilizing-the-low}.

For calculating the direction-averaged conductivity, $\bar{\sigma}=\frac{1}{3}\left(\sigma_{xx}+\sigma_{yy}+\sigma_{zz}\right)$,
we replace the $\xi$ component of the velocity ($\mathcal{V}_{\xi}$)
in the above equations by a random-direction component $\mathcal{V}_{d}=\boldsymbol{\eta}^{T}\mathcal{\boldsymbol{V}}$,
where $\boldsymbol{\eta}=\left(\eta_{x},\eta_{y},\eta_{z}\right)$,
taken as a \emph{random point }on the 3D unit sphere. Since $\mathbb{E}\left[\boldsymbol{\eta}\boldsymbol{\eta}^{T}\right]=\frac{1}{3}\mathcal{I}$,
where $\mathcal{I}$ is the $3\times3$ unit matrix. Averaging over
$\mathcal{V}_{d}$ automatically computes $\bar{\sigma}$.

\begin{figure*}
\begin{centering}
\includegraphics[width=1\columnwidth]{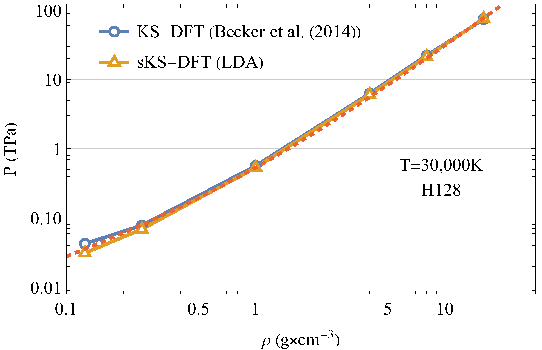}\includegraphics[width=1\columnwidth]{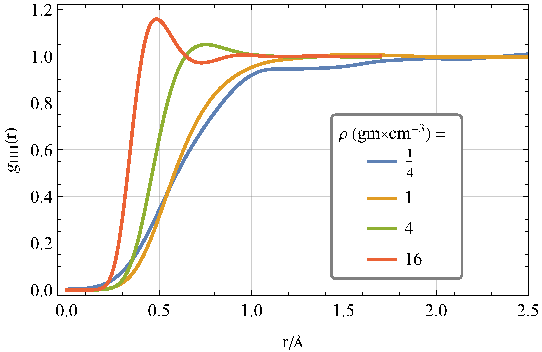} 
\par\end{centering}
\caption{\label{fig:Density-Dependence}\textbf{Left panel}: the estimated
pressure as a function of density for Hydrogen at $T=30,000K$. The
results of the sDFT calculations are shown together with the calculations
taken from ref.~\citep{becker2014abinitio} and a dotted van der
Waals trend of $P\left(\rho\right)=\frac{\rho}{m_{p}}k_{B}T\left(1+\frac{\rho}{\rho_{0}}\right)$,
where $\rho_{0}=0.87\,g/cm^{3}$ is obtained by fitting. \textbf{Right
panel}: the H-H radial distribution for different densities.}
\end{figure*}

\section{\label{sec:Test-case:-Hydrogen}Test case: Hydrogen at 30,000K }

In this section, we use as a test case Hydrogen at 30,000K. We first
consider system-size effects on the estimation of pressure and conductivity.
Then, we give its equation of state, electric conductivity, and radial
distributions, comparing, where possible, with VASP \citep{kresse1993abinitio,kresse1994abinitio,kresse1996efficient,kresse1996efficiency,kresse1999fromultrasoft}.

All sDFT calculations in this section used 40 stochastic orbitals
(for all system sizes), the LDA exchange-correlation energy functional,
and norm-conserving pseudopotentials \citep{troullier1991efficient}.
Unless specifically mentioned otherwise, we use $9\text{E}_{\text{h}}$
cutoff energy and the Baldereschi k-point \citep{baldereschi1973meanvalue}.

\subsection{\label{subsec:System-size-dependent-pressure}System-size dependent
pressure and conductivity }

In Fig.~\ref{fig:Size-Dependence} we present the pressure (left
panel) and the conductivity (right panel) estimates for Hydrogen at
30,000K and density $\rho=1\,\text{g}/\text{cm}^{3}$, as a function
of the system size. The pressure estimates at the Baldereschi k-point
are rather steady and change only mildly with system size. Those done
at the $\Gamma$ point show a stronger sensitivity to system size,
peaking at $H_{256}$ and then decreasing towards the steadier k-point
values. The sDFT pressure estimates increase by about 5\% when going
from cutoff energy of $9$ to $25\text{E}_{\text{h}}$. The dDFT results,
based on the VASP code, are less sensitive to the cutoff energy and
change by only 2.5\%. This reflects the superiority of the PAWs used
by VASP over the norm-conserving pseudopotentials used in the sDFT
calculation when converging to the infinite energy cutoff limit. The
sDFT and the VASP pressure estimates are similar at the higher cutoff
energy.

The size dependence of Hydrogen conductivity at 30,000K, calculated
at the Baldereschi k-point, is shown for two density values in the
right panel of Fig.~\ref{fig:Size-Dependence}. For $1\,\text{g}/\text{cm}^{3}$,
the conductivity curves of $\text{H}_{128}$ and $\text{H}_{512}$
are already quite close (a difference of 3\%). Note that the larger
the system, the smaller the fluctuations. We also show results from
a deterministic calculation on $\text{H}_{128}$, which tend to be
too small as $\omega$ decreases but fit our results well for all
other frequencies.

Finally, the conductivity for the high-density systems is much more
noisy than at low density, and the system size effects are more noticeable,
since the simulation cell size is small for these systems.

\subsection{\label{subsec:Density-dependent-properties-of}Density-dependent
properties of Hydrogen at 30,000K}

The estimated pressure of Hydrogen at $T=30,000K$ as a function of
density is depicted for sDFT and dDFT (using VASP) in the left panel
of Fig.~\ref{fig:Density-Dependence}. The two curves are generally
close. The more significant difference in the lower-pressure estimates
stems primarily from the smaller cutoff energy used in our calculation.
The equation of state can be fitted by a van der Waals form, $P=P_{ideal}\left(1+\frac{\rho}{\rho_{0}}\right)$
with $\rho_{0}=0.87\,\text{g}/\text{cm}^{3}$.

The right panel of Fig.~\ref{fig:Density-Dependence} illustrates
the radial distribution for Hydrogen at $T=30,000$K, with varying
densities from $0.25$ to $16\,\text{g}/\text{cm}^{3}$. At the lowest
density, it reveals a relatively large excluded volume with some corrugated
pattern. As the density increases to $1\,g/cm^{3}$, the radial distribution
curve steepens as the proton-proton repulsion range shortens, and
the corrugated pattern largely dies out, leaving a shallow signature
of a correlation shell at $r=1.5\text{�}$. This feature is enhanced
and contracts to a shorter distance of $0.75\text{�}$ at the density
of $4\,\text{g}/\text{cm}^{3}$. In this regime, the radial distribution
signifies a combined short-range repulsion and longer-range attraction
between pairs, typical of a gas. Finally, at the highest density considered,
$16\,\text{g}/\text{cm}^{3}$, the correlation shell contracts further
to $0.5\text{�}$ while a second correlation shell seems to form $0.9\text{�}$,
hence a radial distribution typical of a liquid emerges at these high
densities.

\begin{figure*}
\centering{}\includegraphics[height=55mm]{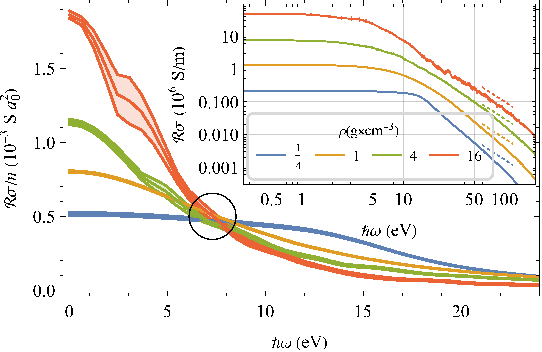}~~~~\includegraphics[height=55mm]{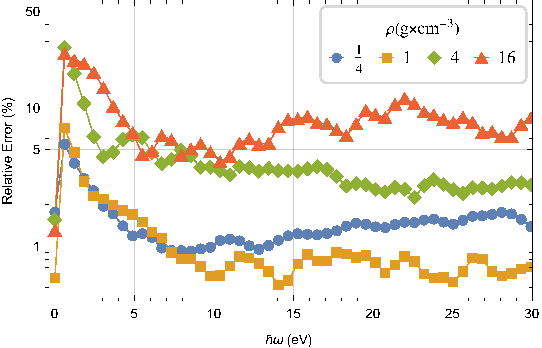}\caption{\label{fig:The-Kubo-Greenwood-estimate}\textbf{Left panel}: The real
part of the Kubo-Greenwood conductivity normalized by the electron
density at $T=30,000K$. The black circle indicates an \emph{isosbestic
point}, which appears at 7.2eV, where all systems have the same normalized
conductivity of $\sim0.5\times10^{-3}\,Sa_{0}^{2}$ . The inset shows
the conductivity in log-log scale, with dashed lines indicating slopes
of $\omega^{-2}$. \textbf{Right panel}: The relative error bars for
the conductivity calculations, as described in the caption of Fig.~\ref{fig:Size-Dependence}.}
\end{figure*}

We now turn to studying hydrogen conductivity using Drude's theory
of metals \citep{ashcroft1976solidstate} as reference point. Drude's
theory gives the real part of the normalized conductivity at frequency
$\omega$ as 
\begin{equation}
\frac{\Re\sigma\left(\omega\right)}{n}=\frac{e^{2}}{m_{e}}\frac{\tau_{c}}{1+\omega^{2}\tau_{c}^{2}},\label{eq:Drude}
\end{equation}
where $n=\frac{N_{e}}{\Omega}$ is the average electron density, and
$\tau_{c},$ the collision time, is the only material parameter. $\tau_{c}$
is assumed independent of $n$. In Fig.~\ref{fig:The-Kubo-Greenwood-estimate}
(left panel) we plot our stochastic estimates of the ab initio normalized
conductivity for Hydrogen in various densities at 30,000K. While the
normalized conductivity in Drude's theory (Eq.~\eqref{eq:Drude})
does not depend on $n$, the ab initio DC normalized conductivity
does depend on it: it changes fourfold as $n$ changes 64fold. Yet,
as seen in Fig.~\ref{fig:The-Kubo-Greenwood-estimate}, at $\hbar\omega=7.2\text{eV}$,
all four \emph{ab initio }curves cross at approximately the same point,
the \emph{isosbestic point}, where they assume \emph{the same }normalized
conductivity value of $0.6\times10^{-3}\text{S}\text{a}_{\text{0}}^{2}$.
The existence of the isosbestic point, especially in the rather large
density range seen here is sometimes indicative of a system composed
of two phases or two states \citep{robinson1999isosbestic,eckstein2007isosbestic,renati2019temperature}.
The value of $\tau_{c}$ at the isosbestic point is equal to $\left(0.084\pm0.035i\right)\,\text{fs}$,
with the real part small relative to typical values of $\tau_{c}$
for room-temperature metals (between 1 and 10 fs \citep{ashcroft1976solidstate}).
This result is consistent with the dense metal (with $r_{s}$ between
0.28 and 0.7) we have, and the high temperatures should speed up relaxation
times and shorten mean free paths in the material. Eq.~\eqref{eq:Drude},
also shows that $\Re\sigma$ is proportional to $\omega^{-2}$ for
high frequencies, $\omega\tau_{c}\gg1$ (in our case, $\hbar\omega>30\text{eV}$).
As seen in the inset of the figure, the ab initio conductivity decreases
faster than the Drude second power law in the frequency.

The relative statistical errors in the conductivity evaluation (before
stabilization according to the method of Appendix~\ref{sec:Stabilizing-the-low})
are shown in the right panel of the figure. In general, they grow
with the density of the gas. Remarkably, the DC conductivity, calculated
through Eq.~\eqref{eq:sigma-DC-OPT 2}, has a much smaller statistical
error than the AC conductivity.

\section{\label{sec:Summary}Summary}

We developed a linear-scaling stochastic DFT implementation in periodic
boundary conditions combined with Langevin dynamics, which we applied
to Hydrogen at $30,000K$. Our pressure estimates at various Hydrogen
densities between $0.125-16\,\text{g}/\text{cm}^{3}$ matched well
with results based on deterministic DFT for high cutoff energy (near
convergence). The sensitivity to the cutoff energy reduced as the
density increased.

The pair correlation functions showed that Hydrogen exhibits gas-like
behavior at densities below $4\,\text{g}/\text{cm}^{3}$ and liquid-like
behavior above it. We also developed a new stochastic method to estimate
the Kubo-Greenwood conductivity with minimal statistical noise at
$\omega\to0$. All the calculations were done in the Baldereschi k-point,
and then the overall size effects in the hydrogen systems were not
large once $H_{512}$ was used, except for the high density which
required a large number of atoms.

Future work will focus on adapting stochastic time-dependent DFT \citep{gao2015sublinear,vlcek2019stochastic,zhang2020linearresponse}
and Green's function methodologies \citep{neuhauser2014breaking,rabani2015timedependent,neuhauser2017stochastic}
for WDM applications, building on the foundation of our current work.

\subsection*{Acknowledgment}

RB and RR thank the German Israel Foundation for funding this project.
ER acknowledges support from the Center for Computational Study of
Excited-State Phenomena in Energy Materials (C2SEPEM) at the Lawrence
Berkeley National Laboratory, funded by the U.S. Department of Energy,
Office of Science, Basic Energy Sciences, Materials Sciences and Engineering
Division, under Contract No. DE-AC02-05CH11231, as part of the Computational
Materials Sciences Program.\medskip{}

\appendix

\section{\label{sec:proof-eq-phi}Proof of Eq.~\eqref{eq:phi_xi_xi'}}

For non-interacting electrons, the Hamiltonian and the grand canonical
distribution are \onecolumngrid
\[
H=\sum_{i}\varepsilon_{i}n_{i},\qquad n_{i}\equiv a_{i}^{\dagger}a_{i},\qquad N=\sum_{i}n_{i},\qquad\rho=\frac{e^{-\beta\left(H-N\mu\right)}}{\text{tr\ensuremath{\left[e^{-\beta\left(H-N\mu\right)}\right]}}}
\]
\twocolumngrid where, $a_{i}$ ($a_{i}^{\dagger})$ are electron
annihilation (creation) operator into eigenstates of the Hamiltonian,
with the standard anti commutation relations $\left\{ a_{i},a_{j}\right\} =\left\{ a_{i}^{\dagger},a_{j}^{\dagger}\right\} =0$,
$\left\{ a_{i}a_{j}^{\dagger}\right\} =\delta_{ij}$.

From these relations alone we find the following: 
\begin{align}
\left[a_{i}^{\dagger}a_{j},a_{k}^{\dagger}a_{l}\right] & =\delta_{jk}a_{i}^{\dagger}a_{l}-\delta_{il}a_{k}^{\dagger}a_{j},\label{eq:comm_one-bode-op}
\end{align}

\begin{equation}
\text{tr}\left[\rho a_{i}^{\dagger}a_{l}\right]=\delta_{il}\text{tr}\left[\rho n_{i}\right]\equiv\delta_{il}p_{\mu\beta}\left(\varepsilon_{i}\right),\label{eq:average_pop:}
\end{equation}
and

\begin{equation}
e^{iHt}a_{k}^{\dagger}a_{l}e^{-iHt}=e^{-i\left(\varepsilon_{l}-\varepsilon_{k}\right)t}a_{k}^{\dagger}a_{l}.\label{eq:evolve-one}
\end{equation}

From Eqs.~\eqref{eq:comm_one-bode-op} and \eqref{eq:average_pop:}
\begin{align}
\text{tr}\left[\rho\left[a_{i}^{\dagger}a_{j},a_{k}^{\dagger}a_{l}\right]\right] & =\left(p_{\mu\beta}^{i}-p_{\mu\beta}^{j}\right)\delta_{jk}\delta_{il}.\label{eq:tr-to-diff-f}
\end{align}
and 
\begin{align}
\text{tr}\left[\rho\left[a_{i}^{\dagger}a_{j},a_{k}^{\dagger}\left(t\right)a_{l}\left(t\right)\right]\right] & =e^{i\left(\varepsilon_{i}-\varepsilon_{j}\right)t/\hbar}\left(p_{\mu\beta}^{i}-p_{\mu\beta}^{j}\right)\delta_{jk}\delta_{il}.\label{eq:td-tr-to-diff-f}
\end{align}
Therefore, first quantization (single-particle) observables $\mathcal{A}$
and $\mathcal{B}$, summed over all electrons $\mathcal{A}=\sum_{n}\mathcal{A}_{n}$
and $\mathcal{B}=\sum_{n}\mathcal{B}_{n}$ correspond, in second quantization,
to $\hat{A}=a_{i}^{\dagger}a_{j}A^{ij}$ and $\hat{B}=a_{i}^{\dagger}a_{j}B^{ij}$
($A^{ij}=\left\langle i\left|\mathcal{A}\right|j\right\rangle $ and
$B^{ij}=\left\langle i\left|\mathcal{B}\right|j\right\rangle $).
We find from Eq.~\eqref{eq:tr-to-diff-f}: 
\begin{align*}
\text{tr}\left[\rho\left[\hat{A},\hat{B}\right]\right] & =\text{Tr}\left[p_{\mu\beta}\left(\mathcal{H}\right)\left[\mathcal{A},\mathcal{B}\right]\right],
\end{align*}
and from these: 
\begin{align}
\Im\,\text{tr}\left[\rho\hat{A}\hat{B}\right] & =\Im\,\text{Tr}\left[p_{\mu\beta}\left(\mathcal{H}\right)\mathcal{A}\mathcal{B}\right].\label{eq:replace-rho-f}
\end{align}
Using Eq.~\eqref{eq:evolve-one} we obtain the generalization of
Eq.~\eqref{eq:replace-rho-f} 
\begin{align}
\Im\,\text{tr}\left[\rho\hat{A}\hat{B}\left(t\right)\right] & =\Im\,\text{Tr}\left[p_{\mu\beta}\left(\mathcal{H}\right)\mathcal{A}\mathcal{B}\left(t\right)\right].\label{eq:gen-replace-rho-f}
\end{align}
This latter equation, used with $\hat{A}\to\sum_{n}\mathcal{R}_{n\xi}$
and $\hat{B}\to\sum_{n}\mathcal{V}_{n\xi'}$ gives Eq.~\eqref{eq:phi_xi_xi'}.

\onecolumngrid

\section{\label{sec:proof-eq-psi}Proof of Eq.~\eqref{eq:sigma-DC-OPT 2}}

The Fourier-transform of Eq.~\eqref{eq:td-tr-to-diff-f} gives: 
\begin{align*}
\int_{-\infty}^{\infty}e^{-i\omega t}\text{tr}\left[\rho\left[a_{i}^{\dagger}a_{j},e^{iHt}a_{k}^{\dagger}a_{l}e^{-iHt}\right]\right]dt & =2\pi\delta\left(\frac{\varepsilon_{i}-\varepsilon_{j}}{\hbar}+\omega\right)\left(p_{\mu\beta}\left(\varepsilon_{j}-\hbar\omega\right)-p_{\mu\beta}\left(\varepsilon_{j}\right)\right)\delta_{jk}\delta_{il},
\end{align*}
where we used the spectral representation of Dirac delta functions,
$2\pi\delta\left(\omega\right)=\int_{-\infty}^{\infty}e^{-i\omega t}dt$,
and the identity $\delta\left(x-y\right)f\left(x\right)=\delta\left(x-y\right)f\left(y\right)$.
Dividing the above expression by $i\omega$ and taking the limit $\omega\to0$
we find: 
\begin{align*}
\lim_{\omega\to0}\frac{1}{i\omega}\int_{-\infty}^{\infty}e^{-i\omega t}\text{tr}\left[\rho\left[a_{i}^{\dagger}a_{j},e^{iHt}a_{k}^{\dagger}a_{l}e^{-iHt}\right]\right]dt & =2\pi\hbar i\delta\left(\frac{\varepsilon_{i}-\varepsilon_{j}}{\hbar}\right)p_{\mu\beta}^{\prime}\left(\varepsilon_{j}\right)\delta_{jk}\delta_{il}.
\end{align*}
Using the spectral representation in the reverse direction, we find:
\[
\lim_{\omega\to0}\frac{1}{i\omega}\int_{-\infty}^{\infty}e^{-i\omega t}\text{tr}\left[\rho\left[a_{i}^{\dagger}a_{j},e^{iHt}a_{k}^{\dagger}a_{l}e^{-iHt}\right]\right]dt=i\hbar\int_{-\infty}^{\infty}e^{-i\left(\varepsilon_{i}-\varepsilon_{j}\right)t}p_{\mu\beta}^{\prime}\left(\varepsilon_{j}\right)dt\delta_{jk}\delta_{il}.
\]
From these, it is now straightforward to show the two one-body observables
\begin{align}
\lim_{\omega\to0}\frac{1}{i\omega}\int_{-\infty}^{\infty}e^{-i\omega t}\text{tr}\left[\rho\left[\hat{A},\hat{B}\left(t\right)\right]\right]dt & =i\hbar\int_{-\infty}^{\infty}\text{Tr}\left[p_{\mu\beta}^{\prime}\left(\mathcal{H}\right)\mathcal{B}\left(t\right)\mathcal{A}\right]dt.\label{eq:full-FT}
\end{align}
For the case $A=B$, the left-hand side can be developed to give an
integral over positive times 
\begin{align}
\lim_{\omega\to0}\frac{1}{i\omega}\int_{-\infty}^{\infty}e^{-i\omega t}\text{tr}\left[\rho\left[\hat{A},\hat{A}\left(t\right)\right]\right]dt & =4i\Re\left[\lim_{\omega\to0}\frac{1}{i\omega}\int_{0}^{\infty}e^{-i\omega t}\Im\,\text{tr}\left[\rho\hat{A}\hat{A}\left(t\right)\right]dt\right].\label{eq:half-FT-left}
\end{align}
The right-hand side of Eq.~\eqref{eq:full-FT} can also be developed
in a similar fashion: 
\begin{align}
i\hbar\int_{-\infty}^{\infty}\text{Tr}\left[p_{\mu\beta}^{\prime}\left(\mathcal{H}\right)\mathcal{A}\left(t\right)\mathcal{A}\right]dt & =2i\hbar\Re\int_{0}^{\infty}\text{Tr}\left[p_{\mu\beta}^{\prime}\left(\mathcal{H}\right)\mathcal{A}\mathcal{A}\left(t\right)\right]dt.\label{eq:half-FT-right}
\end{align}
Equating both right sides: 
\[
\Re\left[\lim_{\omega\to0}\frac{1}{i\omega}\int_{0}^{\infty}e^{-i\omega t}\Im\,\text{tr}\left[\rho\hat{A}\hat{A}\left(t\right)\right]dt\right]=\frac{\hbar}{2}\Re\int_{0}^{\infty}\text{Tr}\left[p_{\mu\beta}^{\prime}\left(\mathcal{H}\right)\mathcal{A}\mathcal{A}\left(t\right)\right]dt.
\]
Finally, using Eq.~\eqref{eq:gen-replace-rho-f} we obtain: 
\begin{equation}
\Re\,\left[\lim_{\omega\to0}\frac{1}{i\omega}\int_{0}^{\infty}e^{-i\omega t}dt\Im\,\text{tr}\left[p_{\mu\beta}\left(\mathcal{H}\right)\mathcal{A}\mathcal{A}\left(t\right)\right]\right]=\frac{\hbar}{2}\int_{0}^{\infty}\Re\,\text{Tr}\left[p_{\mu\beta}^{\prime}\left(\mathcal{H}\right)\mathcal{A}\mathcal{A}\left(t\right)\right]dt,\label{eq:final-proof}
\end{equation}
from which Eq.~\eqref{eq:sigma-DC-OPT 2} can be directly deduced.\medskip{}

\twocolumngrid

\section{\label{sec:Stabilizing-the-low}Stabilizing the low-frequency conductivity
spectrum}

One of the practical problems in calculating the conductivity at low
frequencies arises in connection with introducing a finite resolution
parameter $\nu$ in Eq.~\ref{eq:discrete-integral}. The finite resolution
distorts and usually underestimates the conductivity. This is seen
in the red empty dots of Fig.~\ref{fig:Size-Dependence} which shows
reduced conductivity as $\omega\to0$.

A second problem involves the fact that our conductivity calculations
use a stochastic approach, which has fluctuation errors. For the low-frequency
these fluctuations grow considerably as $\omega\to0$ (see right panel
of Fig.~\ref{fig:The-Kubo-Greenwood-estimate}) due to the division
by $\omega$ in Eq.~\eqref{eq:sigma-AC}.

The two problems described above combine: the high fluctuations in
the low-frequency part of the AC conductivity is exacerbated when
we take smaller $\nu$, needed for converging the AC results to the
DC limit.

Here, we introduce an approximation that helps converge the low frequency
AC conductivity, which relies on the fact that the DC conductivity
($\omega=0$), calculated by a different expression, Eq.~\eqref{eq:sigma-DC-OPT 2},
has finite and small fluctuations (seen in the right panel of Fig.~\ref{fig:The-Kubo-Greenwood-estimate}).
Our stabilization procedure mixes the low-frequency conductivity with
that of an optimized model: 
\[
\sigma_{k}\leftarrow\left(1-w_{k}\right)\sigma_{k}+w_{k}\sigma_{\text{model}}\left(\omega_{k}\right),
\]
where $\sigma_{k}$ is the conductivity corresponding to the frequency
$\omega_{k}=k\nu$, $k=0,1,2,\dots$. Here, $w_{k}$ are mixing weights
\[
w_{k}=\frac{1}{1+\left(\frac{\omega_{k}}{\omega_{c}}\right)^{6}}.
\]
Emphasizing the low-energy spectrum. Since our system exhibits a metallic
behavior, we choose $\omega_{c}$ as the highest frequency in the
spectrum for which $\sigma_{c}>0.7\sigma_{0}$ and we use the Drude
model $\sigma_{\text{model}}\left(\omega\right)=\frac{\sigma_{0}}{1+\omega^{2}\tau_{c}^{2}}$,
depending on the single parameter $\tau_{c}$, the collision time.
Given the calculated spectrum, we set this parameter as follows: 
\[
\tau_{c}^{2}=\frac{\sum_{k}w_{k}\omega_{k}^{2}\sigma_{k}\left(\sigma_{0}-\sigma_{k}\right)}{\sum_{k}w_{k}\omega_{k}^{4}\sigma_{k}^{2}}.
\]
This choice of parameter makes the calculated conductivity values
$\sigma_{k}$ as close as possible, in the root-mean-square sense
to those of the model conductivity $\sigma_{\text{model}}\left(\omega_{k}\right)$,
by minimizing the Lagrangian $L\left(\tau_{c}^{2}\right)=\sum_{k}w_{k}\left(\left(1+\omega_{k}^{2}\tau_{c}^{2}\right)\sigma_{k}-\sigma_{0}\right)^{2}$.

\medskip{}

\bibliographystyle{unsrt}

\end{document}